\documentclass[a4paper,twoside]{article}
\newcommand{\affil}[1]{$^{\rm #1}$}
\usepackage[authoryear]{natbib}
\bibpunct{(}{)}{;}{a}{}{,}
\usepackage{graphicx}
\usepackage{latexsym}
\usepackage{amssymb}
\usepackage{longtable,lscape}
\usepackage{rotating}
\usepackage{longtable}
\usepackage{amsmath}

\date{} %Please leave the date blank
\title{\large\bf\flushleft Medium-resolution {\it s}-process Element Survey of 47 Tuc giant stars}
\author{\parbox{\textwidth}{\flushleft
\vspace{-0.5cm}
{\it C.C. Worley\affil{A,D} and P.L. Cottrell\affil{B,C}}\\
\vspace{0.4cm}
{\small \affil{A}\,Observatoire de la C\^{o}te d'Azur, B.P.4229,
Nice, Cedex 04, France }\\
{\small \affil{B}\,The Beatrice Tinsley Institute, Dept. of Physics
\& Astronomy, University of Canterbury, Private Bag 4800,
Christchurch, New Zealand}\\
{\small \affil{C}\,Max Planck Institut f$\ddot{\textrm{u}}$r Astrophysik, Karl-Schwarzschild Str 1, 85741, Garching, Germany}\\
{\small \affil{D}\,Email: cworley@oca.eu}\\
\vspace{0.2cm}
{\small Accepted for publication in PASA, October 2011}
\vspace{-0.3cm}
}}

\begin{document}
\twocolumn[
\begin{changemargin}{.8cm}{.5cm}
\begin{minipage}{.9\textwidth}
\vspace{-1cm}
\maketitle
%
%
%%%%%%%%%%%%%     ABSTRACT    %%%%%%%%%%%%%
%Abstract of no more than 200 words here.
\small{\bf Abstract:} 

Medium-resolution (R$\sim$6,500) spectra of 97 giant stars in the globular cluster 47 Tucan\ae\ (47 Tuc) have been used to derive the C and N abundance sensitive index, $\delta$C, and to infer abundances of several key elements, Fe, Na, Si, Ca, Zr and Ba for a sample of 13 of these stars with similar $T_{\textrm{eff}}$ and $\log g$.  These stars have stellar properties similar to the well-studied 47 Tuc giant star, Lee 2525, but with a range of CN excess ($\delta$C) values which are a measure of the CN abundance.  The $\delta$C index is shown to be correlated with Na abundance for this sample, confirming previous studies.  The Fe, Ca, Si and the light- and heavy-{\it s} process (slow neutron capture) elements, Zr and Ba respectively, have a narrow range of abundance values in these stars, indicative of a homogeneous abundance within this population of stars.  The constancy of many element abundances (Fe, Si, Ca, Zr, Ba) and the $\delta$C and Na abundance correlation could imply that there has been a second era of star formation in this cluster that has revealed the products of CNO cycle burning via hot bottom burning (depletion of C, enhancement of N and the production of Na for high $\delta$C population). But there is no overall metallicity change across the range of $\delta$C values at a given position in the HR diagram that has been seen in some other globular clusters.

%%%%%%%%%%%%%     KEYWORDS    %%%%%%%%%%%%%
\medskip{\bf Keywords:} globular clusters: individual (47 Tuc) -- stars: abundances -- stars: late-type
% Please write all keywords in lower case. PASA uses the
% standard list of subject headings adopted by The Astrophysical Journal
% and available from http://www.journals.uchicago.edu/ApJ/keywords_text.html.
% Keywords are separated by em-dashes, i.e. ---

%%%%%%%%DO NOT EDIT%%%%%%%%%%%%
\medskip
\medskip
\end{minipage}
\end{changemargin}
]
\small
%%%%%%%%EDIT FROM HERE%%%%%%%%%%%%
\section{Introduction}
Globular clusters (GC) are relics of the birth of the Milky Way galaxy. Their ages, if the order of the age of the Universe, mean that they hold information about early stellar and galactic formation and evolution processes. In particular as closed systems, their chemical evolution can be explored through the chemical signatures of their current stellar components. These signatures are a combination of current nuclear processes within the observed stars and of the material from which these stars formed, and hence of the previous generations of stars that polluted the intra-cluster medium. These signatures are comprised of both light and heavy element abundances, and it is by disentangling these signatures that the pollution and evolution events of GCs can be revealed.

A key goal of our survey of 47 Tuc giant stars was to ascertain the feasibility of determining {\it s}-process element abundances from spectra obtained at the optimal resolution of AAOmega (R$\sim$8,000). These {\it s}-process elements are produced via the slow (compared to $\beta$-decay rate) accumulation of neutrons onto iron seed nuclei that can produce all the elements up to lead. Previous high resolution studies of {\it s}-process elements in 47 Tuc were limited to samples of less than 10 stars. \citet{Worley2010a} and \citet{Wylie2006} respectively observed homogeneities and variations in the {\it s}-process abundance distributions. A statistically significant sample is necessary in order to truly characterise the {\it s}-process abundance distribution of 47 Tuc. This survey was designed to test the resolution limits of abundance determination for weak {\it s}-process features in giant stars.

Our survey was also designed to obtain light element abundances alongside the {\it s}-process elements for a more complete chemical analysis of these stars. There are several key light elemental abundance anomalies that have been observed in globular cluster stars, in particular relating to variations in carbon, nitrogen, oxygen, sodium, magnesium and aluminium. Initially, from low resolution spectra, indices on the CN and CH molecular bands determined an anti-correlation between CH and CN in globular cluster stars \citep{Norris1979}, whereby CN-weak stars are defined generally as having solar C and N abundances, while for CN-strong star N is enhanced by up to $+1$~dex and C is depleted by $\sim -0.4$~dex \citep{Cannon2003}. 

Further study on the CN bimodality showed that while some clusters showed evidence of both CN-weak and CN-strong stars at all stages of evolution \citep[e.g. 47 Tuc,][]{Cannon1998}, others had only one or the other at different stages, although typically both were present on the red giant branch (RGB) \citep[][and references therein]{Gratton2004}. The bimodality was noted to be fairly consistent from the main sequence (MS) to the RGB for any particular globular cluster, but the main discrepancy came in the asymptotic giant branch (AGB) stars where the number of CN-strong stars decreased significantly or there were none at all \citep{Campbell2006,Sneden2000}. It was noted that this could simply be due to a lack of observations of AGB stars, and that increasing the AGB sample to compare to those from the RGB may bear useful results. \citet{Campbell2010} presented preliminary results on such an observational programme which significantly increased the sample sizes of AGB stars for 10 GCs. The preliminary results confirmed that AGB stars in GCs tended to be, in the majority, CN-weak.

Variations in O, Na, Al and Mg have also been observed and appear to be related to the variations in C and N \citep{Cottrell1981}. H-burning via the CNO, NeNa and MgAl cycles affects the abundances of all of these elements. The variations that have been observed are consistent with CNO cycle processing, where Na, Al and Mg are correlated to CN-strength, while O is anti-correlated. However, the observed variations also seem to be correlated with the cluster metallicity. The correlation of Na, Al and Mg with CN-strength is seen in metal-poor clusters, that of Na and Al with CN-strength in less metal-poor clusters and only Na is correlated with CN-strength in metal-rich clusters \citep{Gratton2004}.

These variations in light elements can be attributed to H-burning via CNO cycle during hot bottom burning in intermediate mass AGB stars. Sufficiently high temperatures exist at the bottom of the envelope (top of the H-burning shell) in these stars that allow these events to occur \citep{Cottrell1981}. The CN bimodality in particular is thought to be evidence of a previous generation of CN-weak stars polluting the star forming material such that the CN-strong population are born with material already fully processed in the CNO cycle. The thermally pulsing stage of AGB evolution is also responsible for producing enhancements in light and heavy {\it s}-process elements \citep{Karakas2010}.

\section{Observations}
In October 2008 ninety-seven giant stars in 47 Tuc were observed using AAOmega on the AAT \citep{Sharp2006}. The target stars were selected from: \citet{Lee1977}; \citet{PaltoglouPhD}; \citet{Brown1992}; \citet{Wylie2006}; and \citet{Worley2008}. Photometry was provided by G.~Da Costa and F.~Grundahl (private communication). The co-ordinates of the sample were verified using 2MASS \citep{Skrutskie2006} where $J$ and $K$ magnitudes were also obtained\footnote{http://www.ipac.caltech.edu/2mass}.

\begin{figure}[!h]
\vspace{0.5cm} \centering
\begin{minipage}{75mm}
%\centering
\includegraphics[width = 73mm]{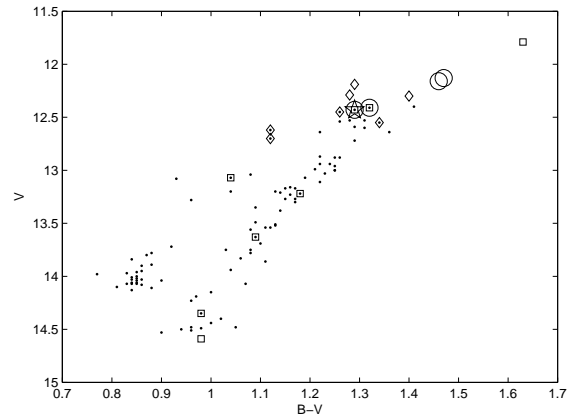}
\caption{Colour-magnitude diagram of the 47 Tuc stars observed in the AAOmega medium resolution survey ($\bullet$). Stars from \citet{Brown1992} ($\bigcirc$), \citet{Wylie2006} ($\lozenge$) and \citet{Worley2008}
($\square$), as well as Lee 2525 (star) are shown.}\label{fig:47tucSurvey_CMD}
\end{minipage}
%\vspace{1cm}
\end{figure}

Figure~\ref{fig:47tucSurvey_CMD} shows the location of the 47 Tuc survey stars in the $V$-($B-V$) colour-magnitude diagram (CMD). The mask configurations for AAOmega restricted stars to within a 2~magnitude range. In order to obtain stars on the tip of the AGB
and on the RGB below the horizontal branch (HB), the range of magnitudes for this survey was $12.5 \leq V \leq 14.5$. Hence there were some stars from either \citet{Brown1992} or \citet{Wylie2006} that could not be observed as they were outside this magnitude range.

\begin{table}[!h]
\centering 
\caption{Wavelength regions and key spectral features observed in the AAOmega 47 Tuc survey.}\label{tab:AAomega_keylines}
\begin{tabular}{cl}
\\
\hline
\footnotesize{Camera:Grating} & \\
\footnotesize{Regions} & Features (\AA) \\

\hline

Blue:3200B &   \\

$\lambda\lambda$ 4040 - 4350 \AA & CN (4142 - 4216) \\

 & CH (4300 - 4330 G Band)\\

\\

Red:2000R &  \\

$\lambda\lambda$ 5750 - 6200 \AA & Na I (6154.23, 6160.75) \\

 & Si I (6131.57, 6131.85, \\

 & \hspace{0.7cm} 6145.02, 6155.13) \\

 & Ca I (6156.02, 6166.44, 6169.56) \\

 & Fe I (6147.83, 6157.73) \\

 & Zr I (6143.18) \\
 & Ba II (6141.73) \\
 & La II (5805.77) \\
 & Nd II (5811.57) \\
\hline
\end{tabular}
\end{table}

AAOmega has a red and a blue camera with which stars can be observed simultaneously over two distinct wavelength ranges. The wavelength regions observed for this survey and the key features that have been analysed are listed in Table~\ref{tab:AAomega_keylines}. The resolution of the spectra was $R\sim 6500$ and the signal-to-noise (SNR) per pixel ranged from 30 to 50 for the blue arm, and 60 to 90 for the red arm. The spectra were reduced using the reduction pipeline, 2dFdr\footnote{http://www.aao.gov.au/2df/manual/UsersManual.pdf}.

The normalisation of the spectra was treated differently for the blue and red arms. The key features in the blue arm observations were CN and CH molecular bands that obscure the location of the true continuum. In the red arm observations molecular features were not so prominent at the stellar temperatures of these stars. In both cases comparison was made to the high resolution spectrum of Arcturus \citep{Hinkle2005} which was convolved to a resolution comparable to the AAOmega observations. Different methods were employed in order to locate the continuum in the blue and the red.

\begin{table}[!b]
\caption{Elemental abundances and abundance ratios with respect to the Sun derived from high (HR) and medium resolution  (MR) spectra of Arcturus using spectrum synthesis.}
\begin{center}
\begin{tabular}{ccccc}
\hline
Species & $\lambda$ (\AA) & $\chi$ (eV) & HR & MR \\ 
\hline
$[$Fe I/H$]$ & 6136.62 & 2.45 & -0.63 & -0.44 \\ 
$[$Fe I/H$]$ & 6147.83 & 4.08 & -0.61 & -0.62 \\ 
$[$Fe I/H$]$ & 6157.73 & 4.08 & -0.60 & -0.52 \\ 
 & \multicolumn{2}{r}{$\langle [$Fe/H$] \rangle$} & -0.61 & -0.53 \\ 
 & \multicolumn{2}{r}{$\sigma$} & 0.02 & 0.09 \\ 
 &  &  &  &  \\ 
$[$Na I/Fe$]$ & 6154.23 & 2.10 & 0.13 & 0.02 \\ 
$[$Na I/Fe$]$ & 6160.75 & 2.10 & 0.19 & 0.10 \\ 
 & \multicolumn{2}{r}{$\langle [$Na/Fe$] \rangle$} & 0.16 & 0.06 \\ 
 & \multicolumn{2}{r}{$\sigma$} & 0.04 & 0.06 \\ 
 &  &  &  &  \\ 
$[$Si I/Fe$]$ & 6131.57 & 5.62 & 0.20 & 0.21 \\ 
$[$Si I/Fe$]$ & 6131.85 & 5.62 & 0.26 & 0.22 \\ 
$[$Si I/Fe$]$ & 6145.02 & 5.62 & 0.24 & 0.18 \\ 
$[$Si I/Fe$]$ & 6155.13 & 5.62 & 0.24 & 0.41 \\ 
 & \multicolumn{2}{r}{$\langle [$Si/Fe$] \rangle$} & 0.24 & 0.26 \\ 
 & \multicolumn{2}{r}{$\sigma$} & 0.03 & 0.10 \\ 
 &  &  &  &  \\ 
$[$Ca I/Fe$]$ & 6156.02 & 2.52 & 0.24 & 0.27 \\ 
$[$Ca I/Fe$]$ & 6166.44 & 2.52 & 0.36 & 0.23 \\ 
 & \multicolumn{2}{r}{$\langle [$Ca/Fe$] \rangle$} & 0.30 & 0.25 \\ 
 & \multicolumn{2}{r}{$\sigma$} & 0.08 & 0.03 \\ 
 &  &  &  &  \\ 
$[$Zr I/Fe$]$ & 6143.18 & 0.07 & 0.02 & -0.12 \\ 
$[$Ba II/Fe$]$ & 6141.73 & 0.70 & -0.20 & 0.04 \\ 
$[$La II/Fe$]$ & 5805.77 & 0.13 & -0.01 & -0.05 \\
$[$Nd II/Fe$]$ & 5811.57 & 0.86 & 0.00 & 0.00 \\
\hline
\end{tabular}
\end{center}
\label{tab:Arcabund_HRLR}
\end{table}

For the red arm, the convolved Arcturus spectrum was compared with synthesised spectra generated using the Arcturus stellar model determined in \citet{Worley2009}. Using the spectrum synthesis programme MOOG \citep{MOOG}, the atomic linelist, collated from the latest laboratory values, was calibrated so that the synthetic spectrum matched the high resolution spectrum for Arcturus. The linelist was modified to include hyper-fine splitting components and isotopic ratios for barium. The abundances for each of the key lines in the red (see Table~\ref{tab:AAomega_keylines}) were measured by spectrum synthesis for both the high- and medium-resolution spectra of Arcturus. Corrections for departures from local thermodynamic equilibrium for the two sodium features were applied after the abundance determination for each star \citep{Lind2011}. The abundances for Arcturus are listed in Table~\ref{tab:Arcabund_HRLR}.

The Ba abundance shows the greatest change between the high-resolution and the convolved Arcutrus spectra. The Ba spectral line that was measured is very strong ($W_{\lambda} \approx 180 \ m$\AA) and so it is sensitive to changes in microturbulence. The remaining elements are in reasonable agreement to $\leq 0.15$~dex between the high-resolution and convolved spectra. These variations in abundance from the high-resolution spectrum to the convolved medium-resolution spectrum provide a measure of the uncertainty in determining the abundances in medium-resolution spectra.

For the blue arm spectral region the high-resolution Arcturus atlas was compared to the convolved Arcturus spectrum in the region of the CN and CH molecular bands. This allowed us to identify pseudo- continuum regions that could be used to normalise the spectra. By enforcing a ratio between the location of three wavelength regions (around 4090\AA, 4220\AA\ and 4318\AA), we created a linearly interpolated profile for each spectrum that was used to undertake the normalisation.  For the spectra of 22 objects, there was no 4318\AA\ region in the spectrum because of the instrument setup and placement of these objects in the focal plane of AAOmega. For these spectra a normalisation based on the normalisation shape for stars of similar $T_{\textrm{eff}}$ was used.

\section{CN indices}\label{sec:AAOmega_CNindex}

CN indices for stars in 47~Tuc have been measured in several studies. A number of stars observed in this survey were previously observed by \citet{Norris1979} and \citet{PaltoglouPhD} (hereafter NF79 and PF84 respectively), in which they were classified by their CN indices, designated here as $\delta$C$_{1979}$. This study seeks to extend these analyses. The CN index used here ($\delta$C$_{2011}$) was that defined by NF79 and PF84:
\begin{align}\label{equ:CNindex}
    S(4142) &= -2.5 log_{10} \Big{\{} \frac{\int ^{4216}_{4120} F_{\lambda}d\lambda }{\int^{4290}_{4216}
    F_{\lambda}d\lambda} \Big{\}}.
\end{align}

NF79 calibrated this line intensity version of the CN index to the previous photometric index, C(4142), using the following equation:
\begin{align}
    C(4142) &= 0.742 \times S(4142) + 0.236.
\end{align}
In order to determine the CN excess, $\delta$C(4142) (hereafter $\delta$C), the CN indices were considered in C(4142)--$V$ space in NF79, and in C(4142)--($B$--$V$) space in PF84. Based on these studies, in this analysis we used the following equation to derive $\delta$C$_{2011}$ for the 47 Tuc sample:
\begin{align}
%  \delta C &=& C(4142) - 0.079 \times V +
%    1.123 ,\label{equ:Norris_CNexcess} \\
  \delta C_{2011} &= C(4142) - (0.304 \times (B-V) - 0.275). \label{equ:Pal_CNexcess}
\end{align}

\begin{figure}[!th]
\centering
\begin{minipage}{75mm}
\includegraphics[width = 70mm]{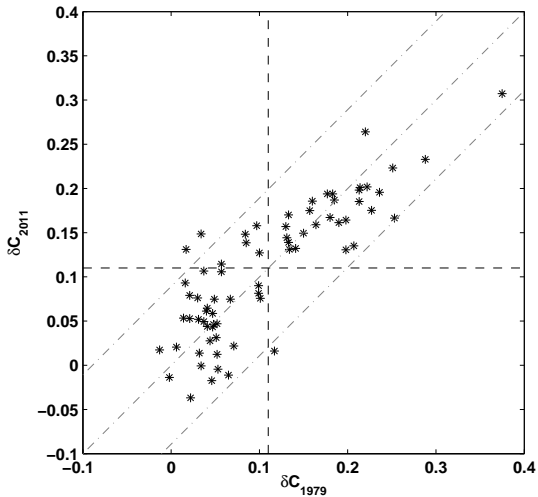} %compCNadj.eps
\caption{Comparison of CN excesses ($\delta$C), from NF79 and PF84 (on abcissa) with the values derived in this study (on ordinate). The dashed lines indicate the CN-weak, CN-strong threshold at $\delta$C~$=0.11$. The dash-dot lines indicate the 1:1 relation $\pm~$2~$\sigma$.}\label{fig:AAOmega_CompCNadj} %\vspace{2cm}
\end{minipage}
\end{figure}

\begin{figure}[!h]
\centering
\begin{minipage}{75mm}
\centering
\includegraphics[width = 70mm]{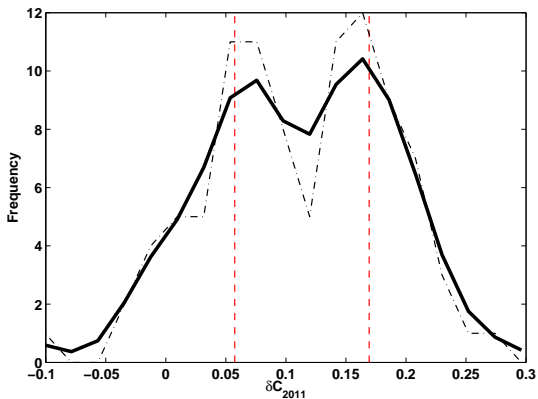}
\caption{Frequency distribution, raw and smoothed, showing the CN bimodality in the 47~Tuc survey using the $\delta$C$_{2011}$ values. The means of each population are shown as red lines.}\label{fig:AAOmega_CNbiHist} %\vspace{2cm} \vspace{-0.5cm}
\end{minipage}
\end{figure}

The stars were classified as CN weak if $\delta$C$_{2011}$~$\leq$~$0.11$ or CN strong if $\delta$C$_{2011}$~$>$~$0.11$ (PF84), as shown in Figure~\ref{fig:AAOmega_CompCNadj}. The photometry, T$_{eff}$ and $\log g$ calculated from $V-K$, and $\delta$C$_{2011}$ values for each star in the survey are given in Appendix~\ref{app:survey}. The majority of the stars fall along the 1:1 relation with a 2~$\sigma$ uncertainty. There are three stars which lie outside the 2~$\sigma$ limits. 

In each case the star has changed designation between CN-strong and CN-weak. Lee 1506 ($\delta$C$_{2011}$ = 0.016) changed from CN-strong to CN-weak, whereas Lee 5703 ($\delta$C$_{2011}$ = 0.131) and Lee 3415 ($\delta$C$_{2011}$ = 0.149) went from CN-weak to CN-strong. The spectra of these outliers were checked for any artifact or normalisation discrepancy but in all three cases the $\delta$C$_{2011}$ value could not be reconciled with the previous measurements. This may indicate misidentification in the previous studies as the coordinates that are used here are consistent with the coordinates of these objects as given in SIMBAD\footnote{http://simbad.u-strasbg.fr/simbad/}.

Figure~\ref{fig:AAOmega_CNbiHist} reduces the CN excess measurements of this study into a histogram using a bin interval of 0.02 (dotted line) and then smoothed using a gaussian filter with a full width at half maximum (FWHM) of 0.05 (solid line) that better reflects the precision of the data (c.f. NF79; PF84). The CN bimodality is seen as two peaks, one for the CN-weak stars at $\delta$C$_{2011}$~$\sim 0.054$ and the other for the CN-strong stars at $\delta$C$_{2011}$~$\sim 0.166$. These values are the mean ($\mu$) values for each population that were determined by the application of a simple gaussian mixture model. The model determined $\sigma = 0.055$ for both populations in mixing proportions of 0.48 for the CN-weak population, and 0.52 for the CN-strong population. The difference in the means, $\Delta\mu = 0.112$, is slightly greater than $2 \sigma = 0.11$ which is the minimum limit necessary to detect two populations in a single dataset\citep{Reschenhofer2001}. Hence two populations in $\delta$C$_{2011}$ exist within this dataset.
 
\begin{figure}[!h]
\begin{minipage}{75mm}
\centering
\includegraphics[width = 80mm]{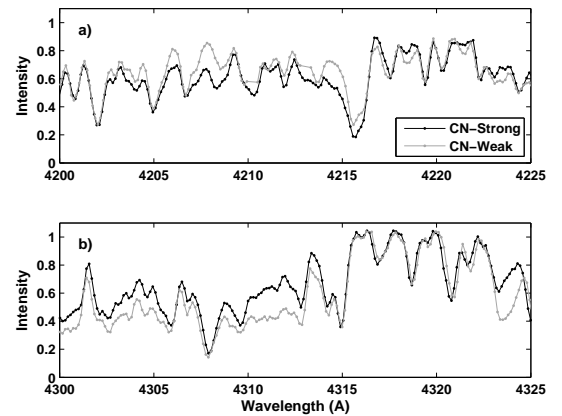}
\caption{Comparison of a CN-weak star (grey, Lee 4506, $\delta$C$_{2011}$=0.050) with a CN-strong star (black, Pal 578, $\delta$C$_{2011}$=0.164) of similar $T_{\textrm{eff}}$ and $\log g$ from the 47~Tuc stellar survey. (a) CN bandhead at 4216~\AA. (b) G Band.}\label{fig:CNwCNs_example}
\end{minipage}
\end{figure}

The CN-CH anti-correlation is a well known feature of 47~Tuc and other globular clusters \citep{Cannon1998}. Figure~\ref{fig:CNwCNs_example} compares a CN-weak and a CN-strong star of similar $T_{\textrm{eff}}$ and {\it log g}. The CN bandhead (degrading to the blue from $\sim$4216~\AA) is distinctly different between the CN-weak and CN-strong stars. At the CH band the CN-strong star has a weak CH band, corresponding to a carbon depletion and nitrogen enrichment from the CN band.

\section{Atmospheric parameters}

\begin{figure}[!th]
\centering
\centering
\includegraphics[width = 82mm]{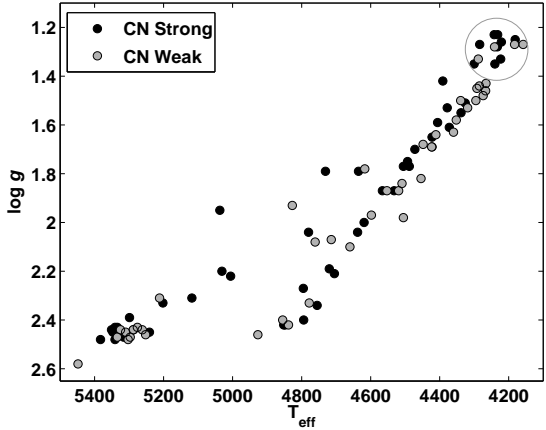}  %TeffLoggCnwCNsVK.eps}
\caption{$T_{\textrm{eff}} - \log g$ space for CN-weak (grey circle) and CN-strong (black circle) pairs based on $V$--$K$ photometry. The circle encompasses Lee~2525-like stars.}\label{fig:CNpairs_Tefflogg}
\end{figure}

The stellar atmospheric models for each star were based on the effective temperature ($T_{\textrm{eff}}$) and surface gravity ($\log g$) values derived from the $V$--$K$ photometry and using the relationships in \cite{Alonso1999}. Comparison between CN-weak and CN-strong stars is ideally investigated using stars at similar $T_{\textrm{eff}}$ and $\log g$. These are shown in $T_{\textrm{eff}} - \log g$ space in Figure~\ref{fig:CNpairs_Tefflogg} and contain many pairs on the RGB, HB and AGB.

The group of thirteen stars (see circle in Figure~\ref{fig:CNpairs_Tefflogg}) at $T_{\textrm{eff}} \approx 4200$~K and {\it log g}~$\approx 1.3$ includes the star Lee~2525, for which a high-resolution analysis was carried out by \cite{Worley2009}. The stars in this group have approximately the same stellar parameters and provide a sample of comparable stars at different CN strengths. The stellar designations, CN excess, photometry and stellar parameters for these stars are listed in Table~\ref{tab:Tenstars_StellarParam}, and are analysed in the remainder of this paper.

\begin{table}[!h]
\centering
\caption{Thirteen stars of similar stellar parameters are listed with their ID, CN excesses taken from NF79 and PF84 ($\delta$C$_{1979}$), and CN excesses calculated in this study ($\delta$C$_{2011}$). The stellar parameters, $T_{\textrm{eff}}$ and {$\log g$}, were calculated for each star based on $V$--$K$ photometry.}\label{tab:Tenstars_StellarParam}
\vspace{0.5cm}
% Table generated by Excel2LaTeX from sheet 'TENL2525Table'
\begin{tabular}{ccccccc}
\hline
\footnotesize{Star ID} & \footnotesize{$\delta$C$_{2011}$} & \footnotesize{$\delta$C$_{1979}$} & Ref. & $T_{\textrm{eff}}$  &  $\log g$ \\ 
\hline
\footnotesize{Pal502}  & -0.04 & 0.02 &  \footnotesize{PF84}  & 4158 & 1.27 \\ 
\footnotesize{Lee2306}  & 0.00 & 0.05 &  \footnotesize{NF79}  & 4288 & 1.33 \\ 
\footnotesize{Lee3622}  & 0.01 & 0.03 &  \footnotesize{NF79}  & 4183 & 1.27 \\ 
\footnotesize{Lee4628}  & 0.04 & 0.05 &  \footnotesize{NF79}  & 4241 & 1.28 \\ 
\footnotesize{Lee2525}  & 0.08 & 0.10 &  \footnotesize{NF79}  & 4232 & 1.23 \\ 
\footnotesize{W66}  & 0.08 &  -  &  -  & 4181 & 1.25 \\ 
\footnotesize{Lee5703} & 0.13 & 0.02 & \footnotesize{NF79} & 4221 & 1.26 \\ 
\footnotesize{Lee1513}  & 0.14 & 0.21 &  \footnotesize{NF79}  & 4242 & 1.23 \\ 
\footnotesize{W139} & 0.15 & - & - & 4223 & 1.33 \\ 
\footnotesize{Pal262}  & 0.17 & 0.25 &  \footnotesize{PF84}  & 4240 & 1.35 \\ 
\footnotesize{W164} & 0.17 & - & - & 4300 & 1.35 \\ 
\footnotesize{Pal661}  & 0.20 & 0.22 &  \footnotesize{PF84}  & 4233 & 1.28 \\ 
\footnotesize{Lee1747}  & 0.22 & 0.25 &  \footnotesize{NF79}  & 4284 & 1.27 \\ 
\\
 &  &  & Mean  & 4233 & 1.28 \\ 
 &  &  & $\sigma$ & $\pm44$ & $\pm0.09$ \\ 
\hline
\end{tabular}
\end{table}

\subsection{Lee 2525}
\begin{figure}[!h]
\centering
\begin{minipage}{75mm}
\includegraphics[width = 75mm]{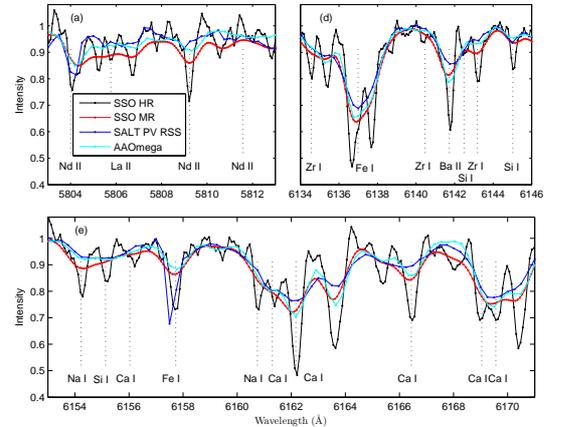}
\caption{Comparison of the Lee~2525 spectra observed using RSS on SALT (blue), the SSO 2.3~m telescope (black), and AAOmega on the AAT (cyan). Also included is the convolution of the high resolution spectra to a resolution comparable to the AAOmega observation (red). (a) Nd II and La II features in the 5800 \AA\ region. (b) Si I, Fe
I, Ba II and Zr I features in the 6140 \AA\ region. (c) Na I, Si I, Ca I and Fe I features in the 6160 \AA\
region.}\label{fig:Lee2525_allSpectra}
\end{minipage}
\end{figure}

Lee 2525 is a 47 Tuc giant star that has been observed in several studies. It has been singled out as a linking star in two previous studies \citep{Wylie2006,Brown1992}. It has been observed in three separate datasets: SALT PV RSS medium resolution observations of eleven stars in 47~Tuc \citep{Worley2008}; AAOmega 47~Tuc medium resolution survey (this study); and high resolution observation of Lee~2525 on the SSO 2.3~m telescope \citep{Worley2009}. Figure~\ref{fig:Lee2525_allSpectra} compares the spectra of Lee~2525 from all three studies in the regions of the light and heavy element spectral features. For comparison, the SSO high resolution spectrum was convolved to a resolution comparable to the AAOmega data in this study.

The Lee~2525 spectrum observed on RSS is at a lower resolution than the AAOmega spectra. The features in common between the high and medium resolution spectra agree in terms of their relative line depths. Abundances derived by spectrum synthesis were obtained for the light and heavy elements for each spectrum, including the convolved SSO spectrum, and are discussed in the next section.

\begin{figure*}[!t]
\centering
\begin{minipage}{150mm}
\begin{minipage}{70mm}
\includegraphics[width = 78mm]{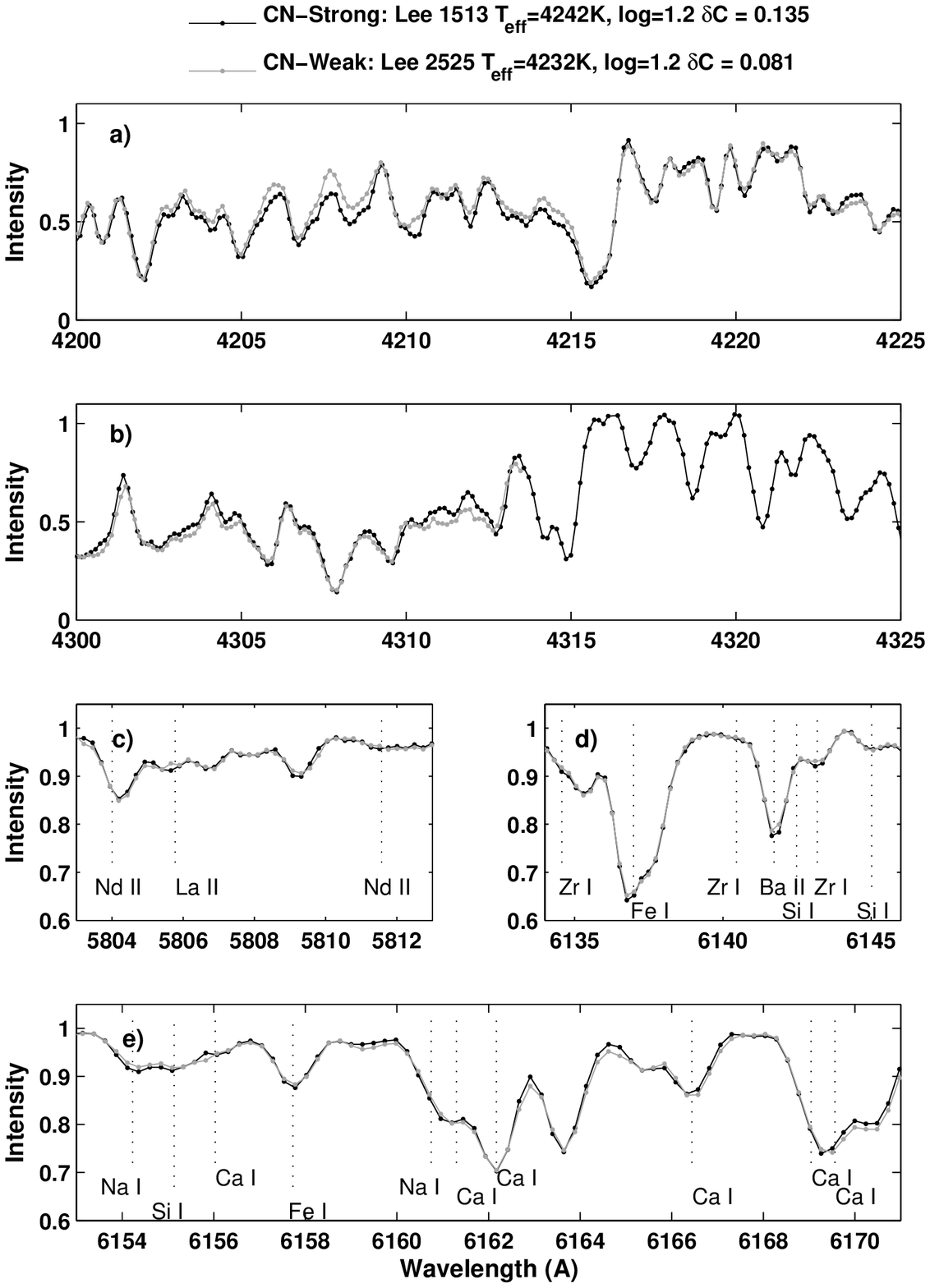}
\caption{Comparison of Lee~2525 (grey) and Lee~1513 (black) spectra. (a) CN band, (b) CH band, (c) Nd II and La II, (d) Si I, Fe I, Ba II and Zr I, (e) Na I, Si I, Ca I and Fe I.}\label{fig:Lee2525L1513spectra}
\end{minipage}
\hfill
\begin{minipage}{70mm}
\includegraphics[width = 78mm]{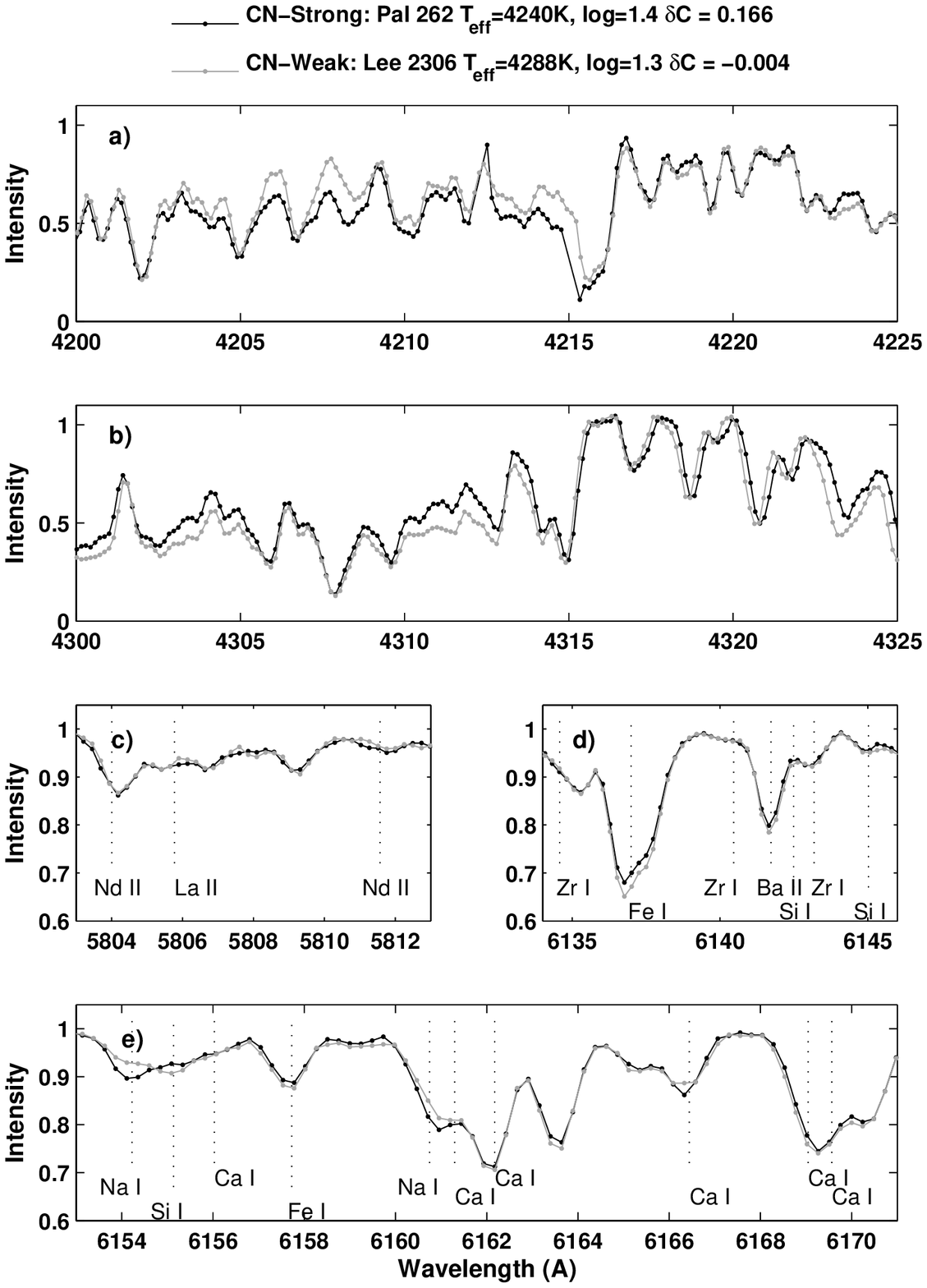}
\caption{As for Figure~\ref{fig:Lee2525L1513spectra}, but comparing the spectra of Pal~262 (black) and Lee~2603 (grey).\vspace{0.9cm}
}\label{fig:Pal262L2603spectra}
\end{minipage}
\end{minipage}
\end{figure*}

In NF79, Lee~2525 was found to be a CN-weak star with a $\delta$C$_{1979}=0.10$. Its CN-strong pair was Lee~1513 which had a CN excess of $\delta$C$_{1979}=0.21$ \citep{Brown1992}. This pairing holds in the current analysis where Lee~2525 was designated as CN-weak ($\delta$C$_{2011}=0.08$) and Lee~1513 as CN-strong ($\delta$C$_{2011}=0.14$), although not as strong as in the previous study.

Figure~\ref{fig:Lee2525L1513spectra} compares the spectra from these two stars, Lee~2525 and Lee~1513. The two spectra are strikingly similar in all regions except that the CN where the smaller $\delta$C in Lee~2525 can be discerned. The remaining spectral regions are also almost an exact match, except that the Na feature at 6154~\AA\ is slightly enhanced in the CN-strong spectrum relative to the CN-weak spectrum.

Figure~\ref{fig:Pal262L2603spectra} compares two stars that are well separated in CN excess value but have similar overall stellar parameters. Pal~262 is a CN-strong star with $\delta$C$_{2011}=0.17$, while Lee~2603 is a CN-weak star with $\delta$C$_{2011}=0.00$. The differences in CN and CH strength are particularly distinct. What also becomes apparent is the difference in the line strength of the two Na I lines at 6154~\AA\ and 6160~\AA. The CN-strong star was considerably stronger in Na I lines than the CN-weak star. The correlation of Na I with CN strength has been noted in previous studies \citep{Gratton2004,Cottrell1981} and will be analysed in Section~\ref{sec:lightelements}. The remaining features, in particular the {\it s}-process element features, show no distinct difference in line strength between these stars.

\section{Lee 2525 sample analysis}
The subset of stars at $T_{\textrm{eff}} \approx 4200$~K and $\log g \approx 1.3$ was used to investigate the abundances of the light and heavy elements in 47~Tuc stars. In the high-resolution analysis of Lee~2525 \citep{Worley2009} the stellar atmospheric model for Lee~2525 was determined to have $T_{\textrm{eff}} = 4225$~K, $\log g = 1.2$, $\xi = 1.8$~kms$^{-1}$ and [Fe/H]~$=-0.70$~dex. This model was used in the following abundance analysis of the medium-resolution observation of Lee~2525. It was subsequently used as the model for all the remaining stars in the subset defined in Table~\ref{tab:Tenstars_StellarParam} as it is within 1~$\sigma$ (10~K and 0.1~dex) of the mean $T_{\textrm{eff}}$ and $\log g$ for all of these stars.

\begin{table*}[!htbp]
\caption{Elemental abundances derived for Lee~2525 from the medium-resolution (MR) AAOmega spectrum, from the spectrum convolved from the high-resolution (C-HR) SSO 2.3~m observation of Lee~2525 and from the high-resolution (HR) analysis itself. The abundances are calculated differentially with respect to the Arcturus MR or HR abundances. Error analysis on the AAOmega medium-resolution spectrum is also included for the specified changes in stellar parameter.}
\begin{center}
\begin{tabular}{cccccccc}
\hline
 & Instrument & AAOmega & SSO 2.3~m & SSO 2.3~m &  &  &  \\ 
 & Resolution & Medium & Convolved & High & \multicolumn{3}{c}{$\Delta[$Fe/H$]$} \\ 
\hline
 &  &  &  &  & $\Delta T_{eff}$ & $\Delta \log g$ & $\Delta \xi$ \\ 
 & $\lambda$ (\AA) & [Fe/H] & [Fe/H] & [Fe/H] & +100~K & 0.5 & 0.5 \\ 
 &  & \multicolumn{3}{l}{---------------------------------------------} & \multicolumn{3}{l}{-----------------------------------} \\ 
Fe I & 6136.62 & -0.13 & -0.03 & 0.08 & 0.03 & 0.10 & -0.33 \\ 
Fe I & 6147.83 & 0.02 & 0.11 & -0.04 & -0.08 & 0.15 & -0.13 \\ 
Fe I & 6157.73 & -0.02 & 0.10 & 0.10 & -0.03 & 0.11 & -0.26 \\ 
 & $\langle [$Fe/H$] \rangle$ & -0.04 & 0.06 & 0.05 & -0.05 & 0.13 & -0.20 \\ 
 & $\sigma$ & 0.08 & 0.08 & 0.08 & 0.13 & 0.16 & 0.21 \\ 
 &  &  &  &  &  &  &  \\ 
X &  & [X/Fe] & [X/Fe] & [X/Fe] & \multicolumn{3}{c}{$\Delta [$X/Fe$]$} \\ 
 &  & \multicolumn{3}{l}{---------------------------------------------} & \multicolumn{3}{l}{-----------------------------------} \\ 
Na I & 6154.23 & 0.00 & 0.23 & 0.12 & 0.06 & -0.02 & -0.09 \\ 
Na I & 6160.75 & -0.08 & -0.10 & 0.15 & 0.08 & 0.00 & -0.12 \\ 
 & $\langle [$Na/Fe$] \rangle$ & -0.04 & 0.06 & 0.14 & 0.07 & -0.01 & -0.11 \\ 
 & $\sigma$ & 0.06 & 0.23 & 0.02 & 0.01 & 0.01 & 0.02 \\ 
 &  &  &  &  &  &  &  \\ 
Si I & 6131.57 & 0.09 & 0.06 & 0.10 & -0.02 & 0.13 & 0.01 \\ 
Si I & 6131.85 & 0.08 & 0.05 & 0.09 & 0.00 & 0.18 & 0.03 \\ 
Si I & 6145.02 & 0.07 & 0.22 & 0.06 & -0.01 & 0.17 & -0.04 \\ 
Si I & 6155.13 & -0.12 & -0.31 & -0.04 & -0.03 & 0.14 & -0.08 \\ 
 & $\langle [$Si/Fe$] \rangle$ & 0.03 & 0.00 & 0.05 & -0.02 & 0.16 & -0.02 \\ 
 & $\sigma$ & 0.01 & 0.22 & 0.06 & 0.03 & 0.03 & 0.07 \\ 
 &  &  &  &  &  &  &  \\ 
Ca I & 6156.02 & -0.14 & -0.21 & -0.14 & 0.05 & 0.03 & -0.07 \\ 
Ca I & 6166.44 & 0.01 & 0.07 & -0.11 & 0.13 & 0.09 & -0.23 \\ 
 & $\langle [$Ca/Fe$] \rangle$ & -0.07 & -0.07 & -0.13 & 0.09 & 0.06 & -0.15 \\ 
 & $\sigma$ & 0.11 & 0.20 & 0.02 & 0.06 & 0.04 & 0.11 \\ 
 &  &  &  &  &  &  &  \\ 
Zr I & 6143.18 & 0.02 & 0.42 & 0.28 & 0.15 & 0.05 & -0.10 \\ 
% &  &  &  &  &  &  &  \\ 
Ba II & 6141.73 & 0.26 & -0.14 & 0.00 & 0.00 & 0.17 & -0.60 \\ 
% &  &  &  &  &  &  &  \\ 
La II & 5805.77 & -0.03 & 0.45 & 0.21 & 0.08 & -0.08 & -0.08 \\ 
% &  &  &  &  &  &  &  \\ 
Nd II & 5811.57 & -0.20 & -0.06 & 0.00 & 0.10 & 0.03 & 0.03 \\ 
\hline
\end{tabular}
\end{center}
\label{tab:AAOmega_LeeAbund}
\end{table*}

In order to make a direct comparison, the high-resolution SSO Lee~2525 spectrum was convolved to the same resolution as the AAOmega spectrum and an abundance analysis was carried out. The results from that analysis and a spectrum synthesis analysis of the high-resolution Lee~2525 spectrum are also listed in Table~\ref{tab:AAOmega_LeeAbund}\footnote{From this section onwards the definition of the bracket notation [X/H] or [X/Fe] is $\log_{10}$(X/H)$_{\bigstar}- \log_{10}$(X/H)$_{\textrm{Arcturus}}$ or $\log_{10}$(X/Fe)$_{\bigstar}- \log_{10}$(X/Fe)$_{\textrm{Arcturus}}$}.

\begin{table*}[!th]
\vspace{1.5cm}
\caption{Elemental abundances derived for thirteen stars in the 47~Tuc medium-resolution survey with stellar parameters similar to Lee~2525. The abundances determined for the convolved high-resolution (C-HR) Arcturus atlas are also listed and are with respect to the Sun, whereas the 47~Tuc abundances are calculated differentially with respect to these Arcturus values. The Lee~2525 stellar atmospheric model was used to infer the abundances for each of these stars. The mean abundances and standard deviations for the entire  47 Tuc sample are also listed.}
\begin{center}
\begin{tabular}{rccccccccc}
\hline
Star & {\footnotesize Arcturus} & {\footnotesize Pal502} & {\footnotesize Lee2306} & {\footnotesize Lee3622} & {\footnotesize Lee4628} & {\footnotesize Lee2525} & {\footnotesize W66} & {\footnotesize Lee5703} & {\footnotesize Lee1513} \\ 
\hline %&  &  &  &  &  &  &  &  &  \\ 
%$T_{\textrm{eff}}$ & 4300 & 4158 & 4288 & 4183 & 4241 & 4232 & 4181 & 4221 & 4242 \\ 
%$\log g$ & 1.6 & 1.27 & 1.33 & 1.27 & 1.28 & 1.23 & 1.25 & 1.26 & 1.23 \\ 
$\xi$ & 1.5 & 1.8 & 1.8 & 1.5 & 1.8 & 1.8 & 1.8 & 1.8 & 2.0 \\ 
$\delta$C$_{2011}$ & -0.06 & -0.04 & 0.00 & 0.01 & 0.04 & 0.08 & 0.08 & 0.13 & 0.14 \\ 
$[$Fe/H$]$ & -0.53 & -0.20 & -0.14 & -0.14 & -0.13 & -0.14 & -0.15 & -0.14 & -0.15 \\ 
$[$Na/Fe$]$ & 0.06 & -0.06 & -0.21 & -0.13 & -0.18 & -0.10 & 0.06 & -0.06 & -0.02 \\ 
$[$Si/Fe$]$ & 0.26 & -0.06 & 0.02 & -0.08 & 0.03 & 0.03 & -0.08 & -0.02 & 0.01 \\ 
$[$Ca/Fe$]$ & 0.25 & 0.05 & -0.14 & 0.03 & -0.03 & -0.07 & -0.01 & -0.07 & -0.12 \\ 
$[$Zr/Fe$]$ & -0.12 & 0.32 & 0.07 & 0.22 & 0.22 & 0.02 & 0.27 & 0.09 & 0.10 \\ 
$[$Ba/Fe$]$ & 0.04 & 0.21 & 0.24 & 0.29 & 0.31 & 0.26 & 0.21 & 0.36 & 0.29 \\ 
$[hs/ls]$ & 0.16 & -0.11 & 0.17 & 0.07 & 0.09 & 0.24 & -0.06 & 0.27 & 0.19 \\ 
\hline &  &  &  &  &  &  &  &  &  \\ 
 &  &  &  &  &  &  &  &  &  \\ 
Star & {\footnotesize W139} & {\footnotesize Pal262} & {\footnotesize W164} & {\footnotesize Pal661} & {\footnotesize Lee1747} & \multicolumn{1}{l}{} & \multicolumn{1}{l}{} & \multicolumn{ 2}{c}{Sample Statistics} \\ 
\hline %&  &  &  &  &  &  &  &  &  \\ 
%$T_{\textrm{eff}}$ & 4223 & 4240 & 4300 & 4233 & 4284 & \multicolumn{ 2}{r}{$\langle T_{eff} \rangle$} & 4233 & 46 \\ 
%$\log g$ & 1.33 & 1.35 & 1.35 & 1.28 & 1.27 & \multicolumn{ 2}{r}{$\langle \log g \rangle$} & 1.28 & 0.09 \\ 
$\xi$ & 1.8 & 1.5 & 1.8 & 1.8 & 1.8 & \multicolumn{ 2}{r}{$\langle \xi \rangle$} & 1.8 & 0.13 \\ 
$\delta$C$_{2011}$ & 0.15 & 0.17 & 0.17 & 0.20 & 0.22 & \multicolumn{ 2}{r}{$\langle \delta$C$_{2011}$~$\rangle$} & 0.10 & 0.09 \\ 
$[$Fe/H$]$ & -0.13 & -0.16 & -0.16 & -0.12 & -0.11 & \multicolumn{ 2}{r}{$\langle$[Fe/H]$\rangle$} & -0.14 & 0.02 \\ 
$[$Na/Fe$]$ & -0.08 & 0.25 & 0.02 & 0.02 & 0.17 & \multicolumn{ 2}{r}{$\langle$[Na/Fe]$\rangle$} & -0.02 & 0.13 \\ 
$[$Si/Fe$]$ & 0.04 & -0.01 & -0.08 & 0.00 & 0.01 & \multicolumn{ 2}{r}{$\langle$[Si/Fe]$\rangle$} & -0.01 & 0.04 \\ 
$[$Ca/Fe$]$ & -0.14 & -0.03 & -0.11 & -0.09 & -0.12 & \multicolumn{ 2}{r}{$\langle$[Ca/Fe]$\rangle$} & -0.06 & 0.06 \\ 
$[$Zr/Fe$]$ & 0.12 & 0.12 & 0.12 & 0.07 & 0.07 & \multicolumn{ 2}{r}{$\langle$[Zr/Fe]$\rangle$} & 0.14 & 0.09 \\ 
$[$Ba/Fe$]$ & 0.26 & 0.31 & 0.36 & 0.26 & 0.26 & \multicolumn{ 2}{r}{$\langle$[Ba/Fe]$\rangle$} & 0.28 & 0.05 \\ 
$[hs/ls]$ & 0.14 & 0.19 & 0.24 & 0.19 & 0.19 & \multicolumn{ 2}{r}{$\langle [hs/ls] \rangle$} & 0.14 & 0.11 \\
\hline
\end{tabular}
\end{center}
\label{tab:AAOmega_AllAbund}
%\vspace{0.5cm}
\end{table*}

There is reasonable agreement in the abundance for Fe for all three sets of results with a similar level of uncertainty ($\sim$0.08~dex). For the light elements, the convolved SSO spectrum produced the largest spread in values ($\sim$0.2~dex) between the lines for each of Na, Si and Ca, while for the other two spectra there were smaller uncertainties in each ($\leq$~0.11~dex). Clearly the information retained in a simple gaussian convolution does not necessarily match observations at the resolution to which the convolution was made. 

For each of the light and heavy elements there are discrepencies between the abundance determinations of these different Lee~2525 spectra which is an example of how the degradation of spectra to lower resolution imply a different abundance of an element. Also the high resolution spectrum of Lee~2525 had SNR$\sim$50 so features that are due to the noise are likely to have contaminated the shape of the spectral features in the convolution. There was much better agreement between abundances derived from the Arcturus high resolution and convolved spectra (see Table~\ref{tab:Arcabund_HRLR}). A high SNR, high resolution observation of Lee~2525 would provide the basis for a more consistent comparative analysis.

The error analysis in Table~\ref{tab:AAOmega_LeeAbund} shows changes in the elemental abundances on the order of 0.1~dex for changes in the stellar parameters based on the intervals in the stellar model grid.

With regard to the {\it s}-process element abundances, the spectral features for La and Nd are heavily blended at this resolution (see Figure~\ref{fig:Lee2525_allSpectra}) and only the Ba line at 6141.73~\AA\ and the Zr line at 6143.18~\AA\ looked sufficiently distinct for analysis. Abundances were determined for La and Nd for the Lee 2525 spectra, but the features were deemed to be too blended ($R\sim 6500$) for analysis in the remaining survey stars.

\begin{figure*}[!th]
\centering
\begin{minipage}{160mm}
%\hspace{-1.0cm}
%\centering
%\includegraphics[width = 175mm]{AAOmega/zrbahslsVfe.eps}
\begin{minipage}{50mm}
\includegraphics[width = 53mm]{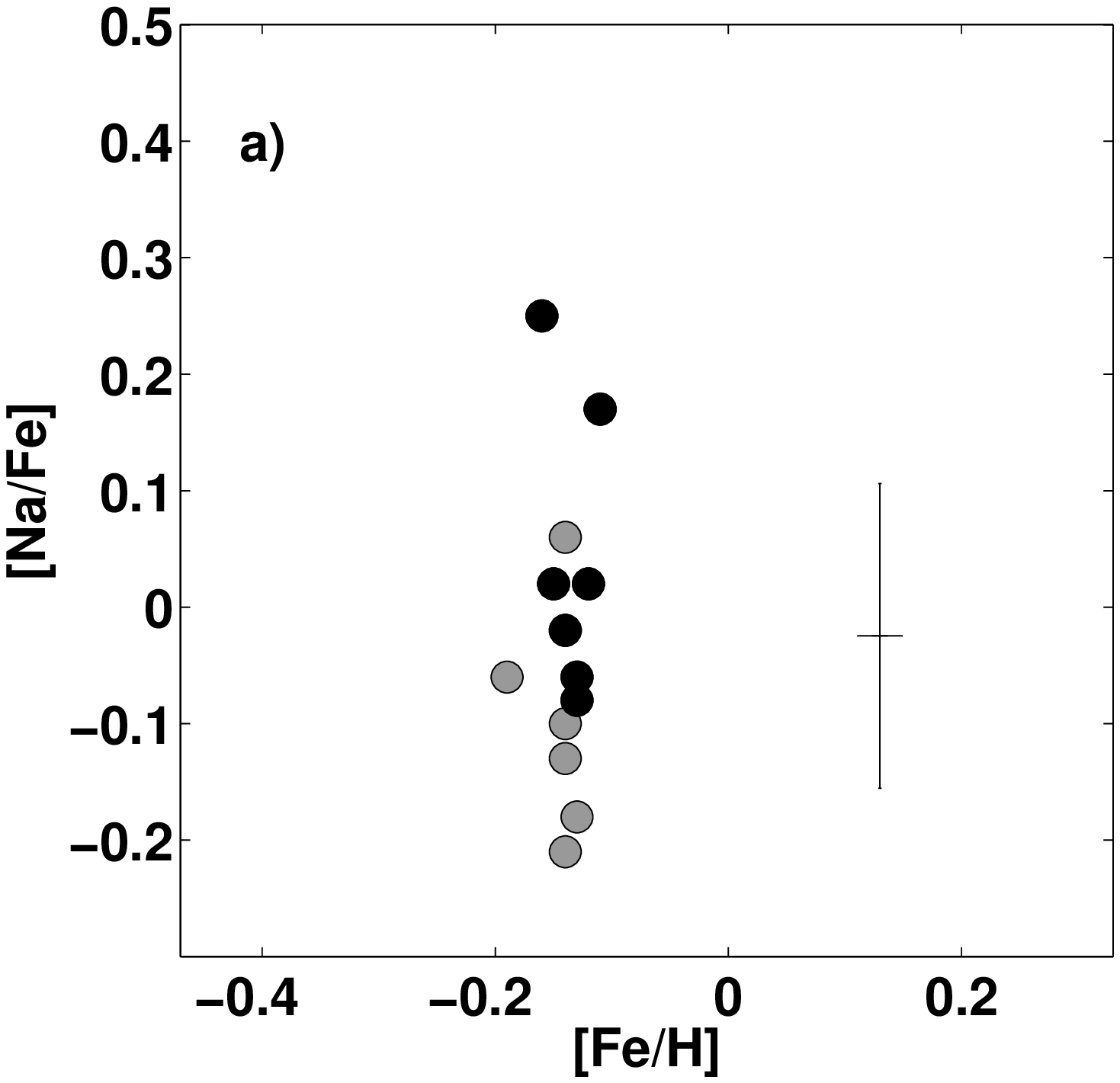}  %nasicaVfe.eps
\end{minipage}
\hspace{0.1cm}
\begin{minipage}{50mm}
\includegraphics[width = 53mm]{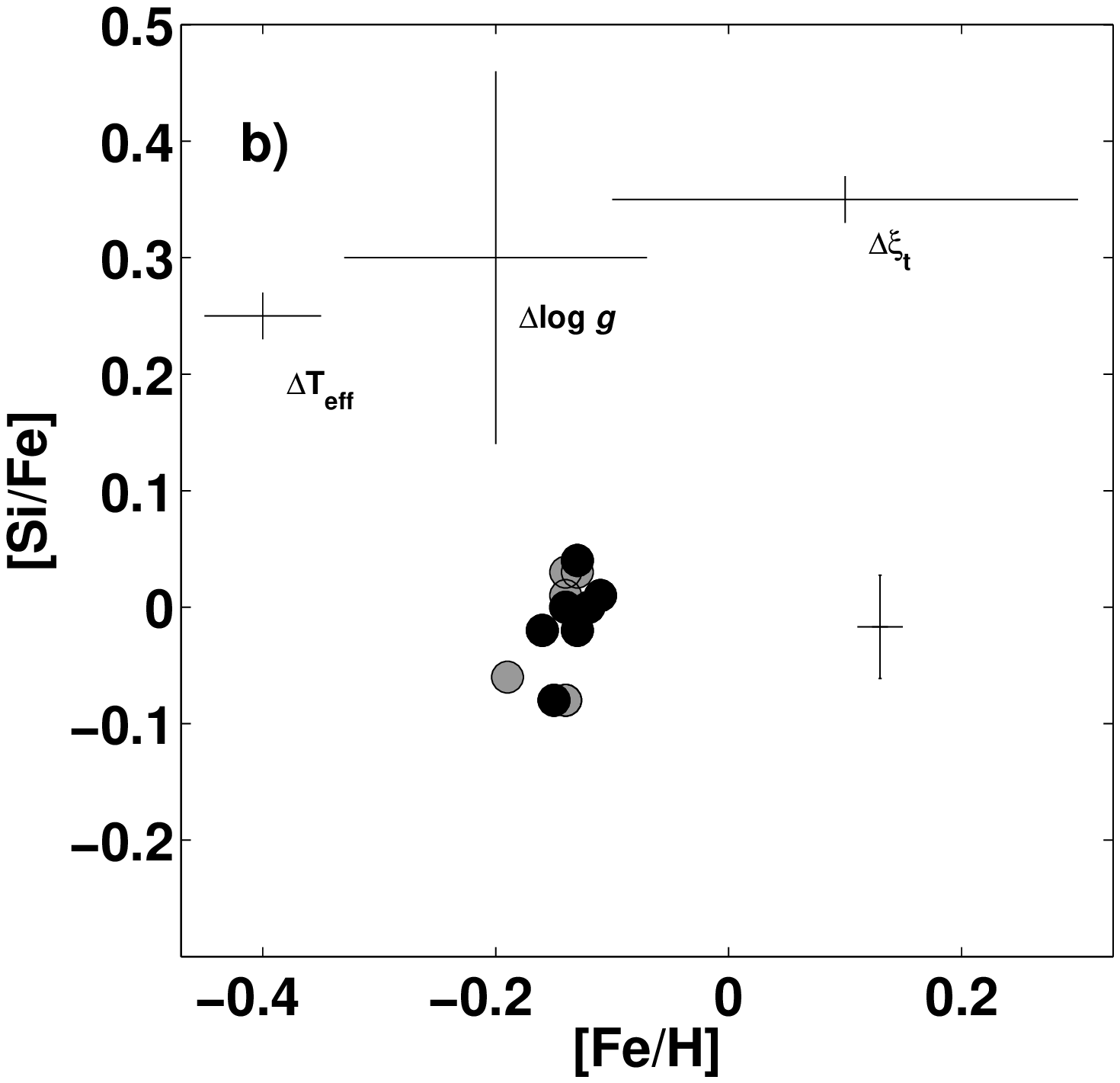}  %nasicaVfe.eps
\end{minipage}
\hspace{0.1cm}
\begin{minipage}{50mm}
\includegraphics[width = 53mm]{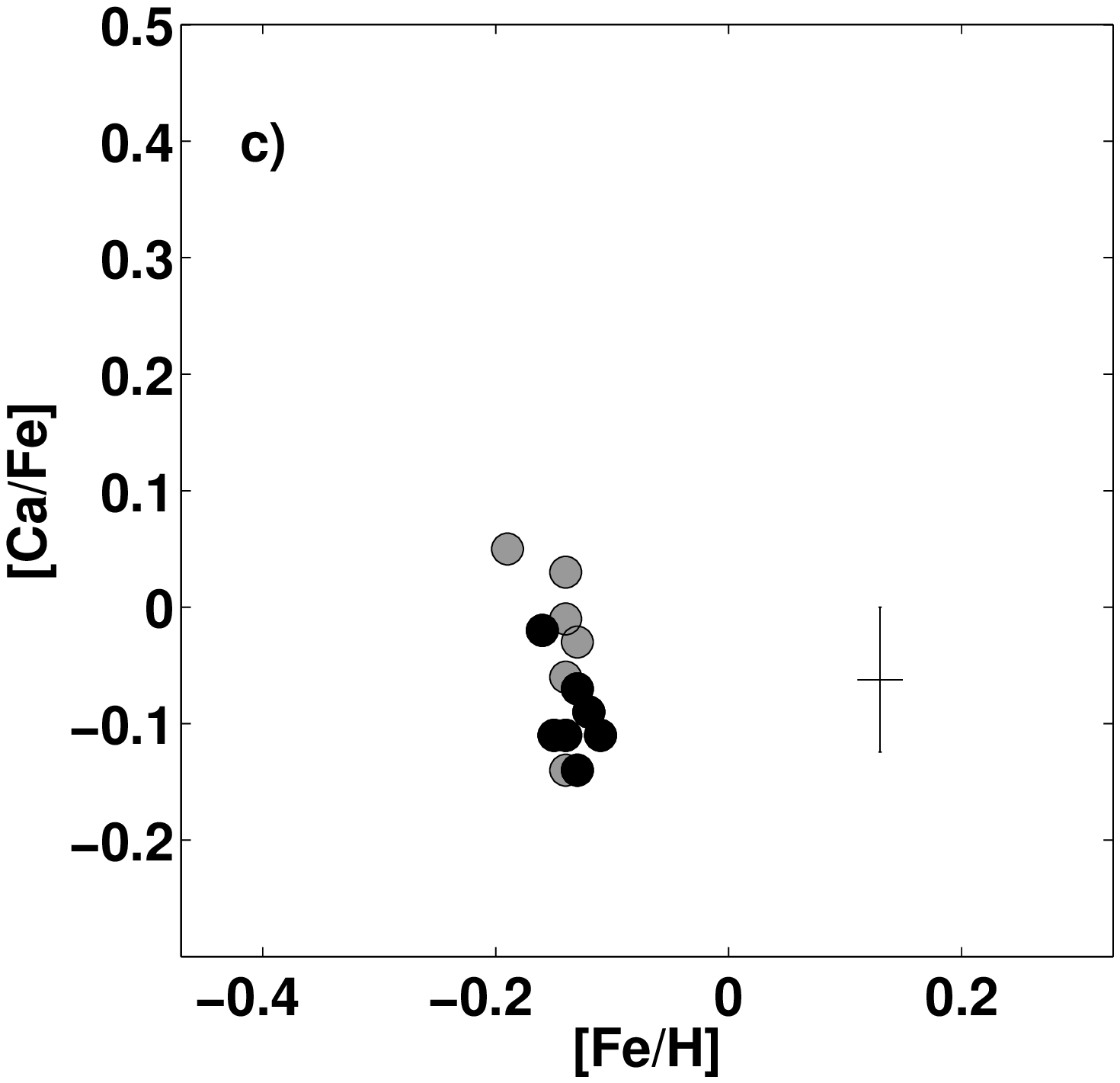}  %nasicaVfe.eps
\end{minipage}
\caption{The element abundance ratios [X/Fe] calculated with respect to the medium resolution abundance analysis of Arcturus for a) Na, b) Si and c) Ca against [Fe/H] for each of the thirteen `Lee~2525-like' stars. Grey circles are CN-weak stars and black circles are CN-strong. The sample mean and standard deviation for each is also shown. The systematic uncertainties from Table~\ref{tab:AAOmega_LeeAbund} are illustrated for [Si/Fe].}\label{fig:TenStars_lightVFe}
%\vspace{1.5cm}
\end{minipage}
\end{figure*}

Following this analysis of the medium-resolution Lee~2525 spectra the other stars in Table~\ref{tab:Tenstars_StellarParam} were analysed for their elemental abundances. As noted earlier, the model determined for Lee~2525 was used in the analysis of the spectra for each of these stars. The only parameter that was varied was the microturbulence ($\xi_t$) as it was clear for four of the stars that the Lee~2525 microturbulence value was not a good fit. Table~\ref{tab:AAOmega_AllAbund} lists the stellar parameters, CN excess and elemental abundances for each of the thirteen stars in this subset. The [{\it hs/ls}] ratio is included as the difference between the Ba (heavy {\it s}-process) and Zr (light {\it s}-process) abundances.

\subsection{Light elemental abundances}\label{sec:lightelements}
The light elemental abundance ratios with respect to Arcturus are compared with [Fe/H] in Figure~\ref{fig:TenStars_lightVFe}. The [Si/Fe] and [Ca/Fe] abundance ratios each have a small spread  (of $<$~0.2~dex) for this sample, indicating that there is a homogeneous abundance distribution for these two elements. The systematic uncertainties with associated changes in $T_{\textrm{eff}}$, $\log g$ and $\xi_{t}$ presented in Table~\ref{tab:AAOmega_LeeAbund} are shown for [Si/Fe] and [Fe/H] in Figure~\ref{fig:TenStars_lightVFe}b to illustrate the range in values with changes in stellar parameters. The spread in the [Na/Fe] far exceeds what can be expected from the corresponding systematic uncertainties of $\sim$~0.2~dex (Table~\ref{tab:AAOmega_LeeAbund}). This implies that the range in [Na/Fe] is real.

\begin{figure}[!h]
\centering
\begin{minipage}{75mm}
\centering
\includegraphics[width = 75mm]{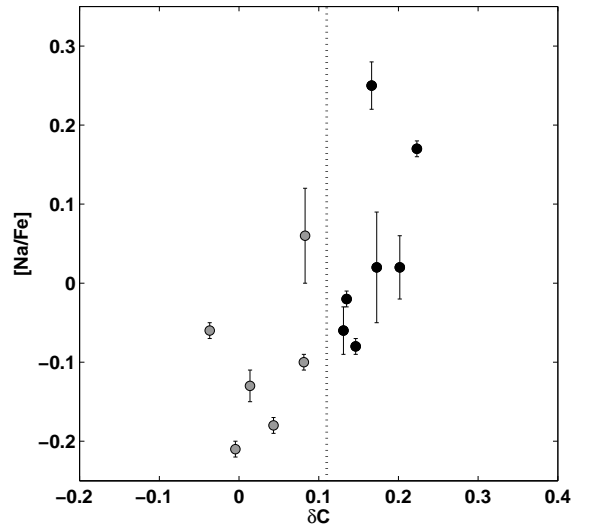}
\vspace{-0.4cm}\caption{[Na/Fe] with respect to Arcturus against $\delta$C$_{2011}$ for each of the thirteen Lee~2525-like stars. The dotted line shows the threshold at 0.11. Grey circles are CN-weak stars and black circles are CN-strong.} \label{fig:TenStars_NavdC}
\end{minipage}
\end{figure}

Furthermore the Na-CN strength correlation is well documented in globular cluster stars \citep{Cottrell1981}. Figure~\ref{fig:TenStars_NavdC} compares the [Na/Fe] abundance ratio with CN strength ($\delta C_{2011}$). There is a clear correlation, with the increasing enrichment in Na corresponding to increasing CN strength. The trend seems continuous rather than attributable to two distinct populations. Of the stars in the Lee~2525-like sample 10 have previously determined $\delta$C values, one of which is Lee~5703 for which the $\delta$C$_{2011}$ differed above 2~$\sigma$ from the $\delta$C$_{1979}$ value. However is it not one of the outliers in the Figure~\ref{fig:TenStars_NavdC}, nor for the remaining abundance determinations, implying that the current analysis of that star is reliable.

\subsection{Heavy elemental abundances}
\begin{figure*}[!th]
\centering
\begin{minipage}{160mm}
%\hspace{-1.0cm}
%\centering
%\includegraphics[width = 175mm]{AAOmega/zrbahslsVfe.eps}
\begin{minipage}{50mm}
\includegraphics[width = 53mm]{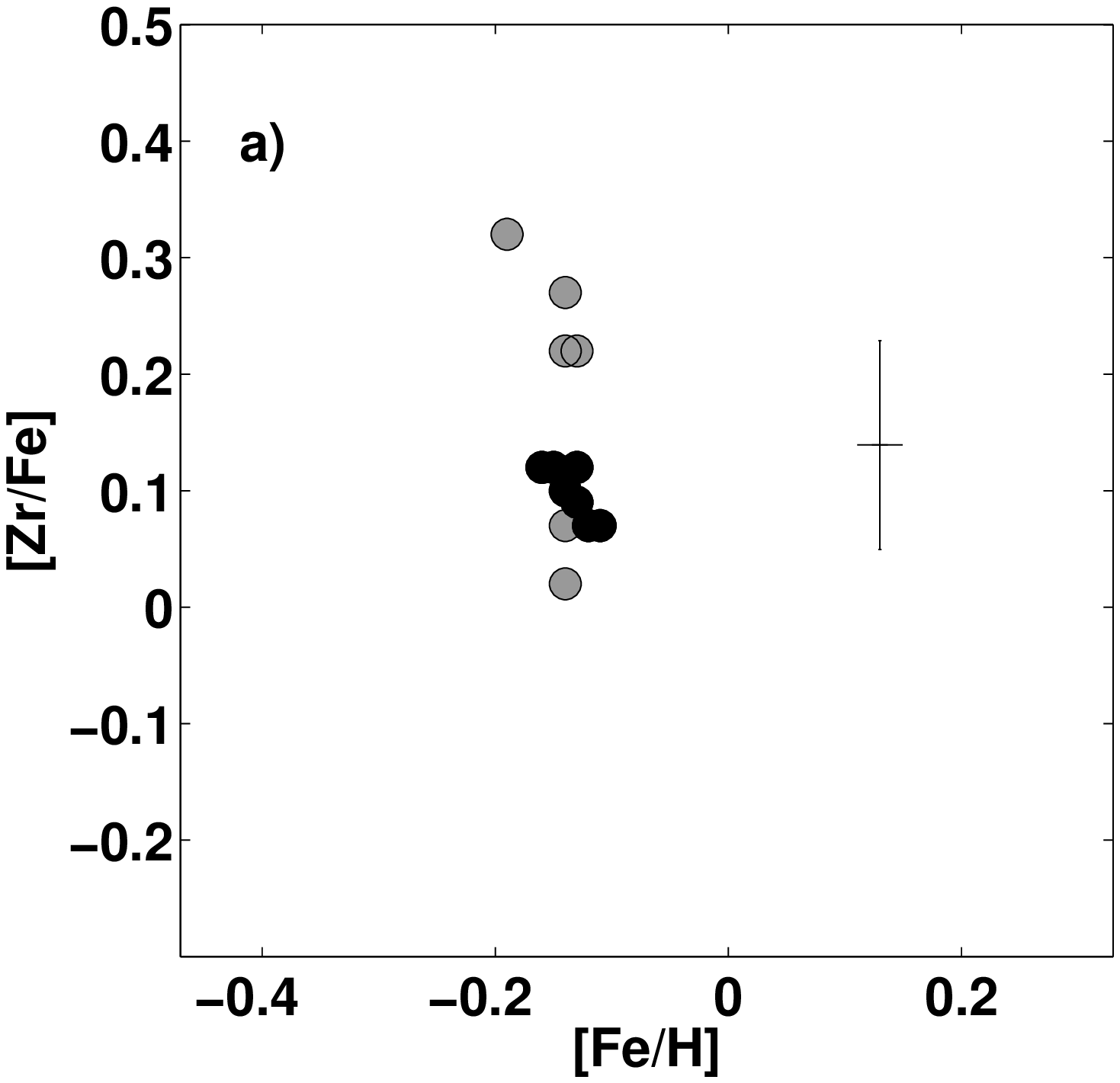}  %nasicaVfe.eps
\end{minipage}
\hspace{0.1cm}
\begin{minipage}{50mm}
\includegraphics[width = 53mm]{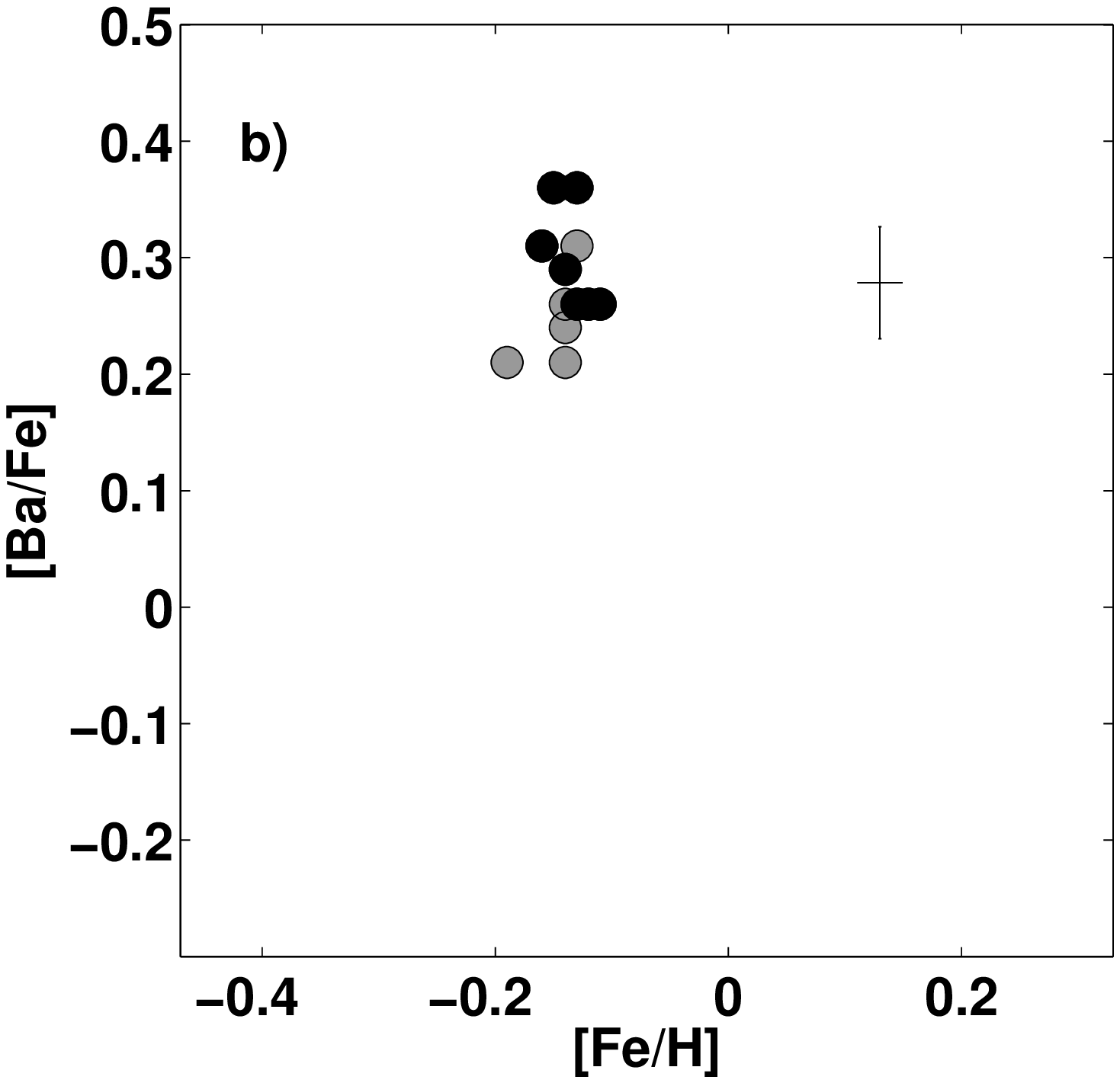}  %nasicaVfe.eps
\end{minipage}
\hspace{0.1cm}
\begin{minipage}{50mm}
\includegraphics[width = 53mm]{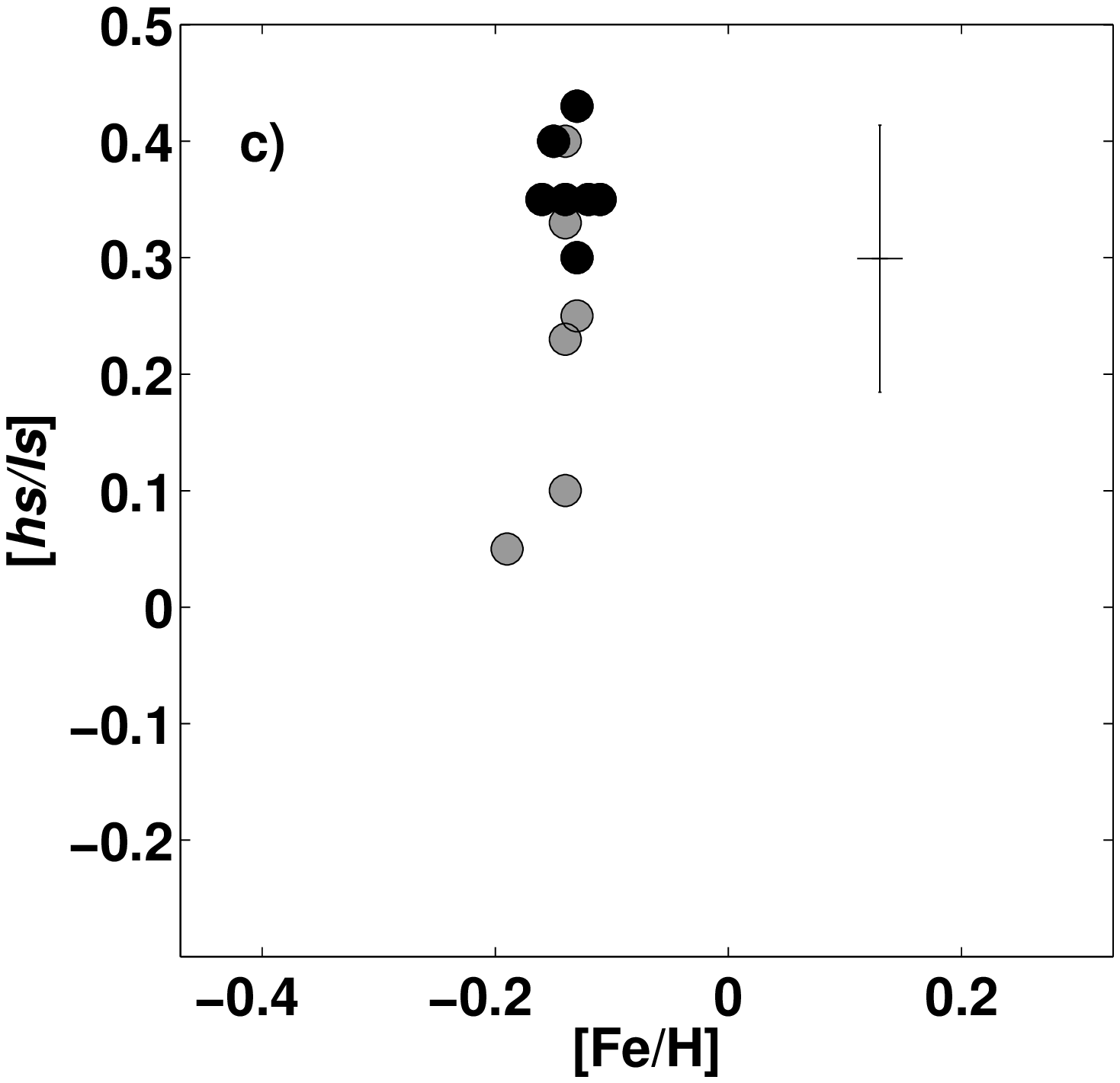}  %nasicaVfe.eps
\end{minipage}
\caption{The abundances of the {\it s}-process elements with respect to Arcturus, a) Zr and b) Ba, against [Fe/H] for the thirteen Lee~2525-like stars in the 47~Tuc medium-resolution survey with sample mean and standard deviation. c) The ratio of [$hs/ls$] against [Fe/H] for the sample. Grey circles are CN-weak stars and black circles are CN-strong.}\label{fig:TenStars_zrbaVfe}
%\vspace{1cm}
\end{minipage}
\end{figure*}

Figures~\ref{fig:TenStars_zrbaVfe}a and b show the abundance ratios for each of [Zr/Fe] and [Ba/Fe] against [Fe/H] for the Lee~2525-like sample. In Figure~\ref{fig:TenStars_zrbaVfe}a there is a distinct spread in the values for [Zr/Fe]. However the Zr feature  used in this analysis (see Figure~\ref{fig:Lee2525L1513spectra}) is small and quite blended at this resolution so the larger uncertainly is not unexpected. There is however a very small spread in the [Ba/Fe] abundance ratios (Figure~\ref{fig:TenStars_zrbaVfe}b) that is similar to that for [Si/Fe] in Figure~\ref{fig:TenStars_lightVFe}b. The Ba feature used here is very senstive to $\xi_t$. However the small spread indicates that the adjustment of the $\xi_t$ values resulted in consistency between the Ba and Fe abundances.

Figure~\ref{fig:TenStars_zrbaVfe}c shows the distribution of the [$hs/ls$] ratio against [Fe/H] for this sample. The spread in [Fe/H] is very small at $0.02$~dex while the larger spread in the [$hs/ls$] values can mainly be attributed to the spread in the Zr abundances, hence they are subject to large systematic uncertainties. The sample mean value of $\langle$[$hs/ls$]$\rangle$~=~0.14$\pm$0.11~dex is slightly higher than some recent analysis of high resolution spectra of 47 Tuc giant stars, [$hs/ls] = -0.13 \pm 0.05$~dex \citep{Worley2010a}. Both of these values are indicative of values for stars at this metallicity \citep{Busso2001}. This analysis shows that to properly survey the weak {\it s}-process spectral features a higher resolution is needed.

\section{Conclusion}
The medium-resolution survey of 47~Tuc stars has provided a good sample with which to investigate several aspects of chemical abundances in GCs. The survey sample clearly shows the CN-CH anti-correlation which is well-documented for GCs \citep{Norris1979, Cannon1998, Cannon2003, Briley2004}. The bimodal distribution of CN-weak and CN-strong stars is also evident in the sample. The current measurement of the CN indices of these stars were in reasonable agreement with the previous studies of NF79 and PF84.

The grouping of stars about Lee~2525 provided a unique sample with which to investigate abundance variations for stars of very similar stellar parameters. To complement the CN indices, the medium-resolution spectra were analysed for light and heavy element abundnaces using spectrum synthesis techniques. This preliminary set of stars, and their representative star, Lee~2525, provided a link to high-resolution abundance analyses of 47 Tuc giant stars in previous studies \citep{Wylie2006,Brown1992,Worley2008,Worley2010a}.

Relative to Arcturus, the abundance analyses for these stars give a strong indication for a homogenous distribution  of Fe ($\langle$[Fe/H]$\rangle =-0.14\pm0.02$~dex), Si ($\langle$[Si/Fe]$\rangle =-0.01\pm0.04$~dex) and Ca ($\langle$[Ca/Fe]$\rangle =-0.06\pm0.06$~dex) in 47~Tuc. There is a much larger scatter in the abundance derived for Na ($\langle$[Na/Fe]$\rangle =-0.05\pm0.14$~dex) and the Na abundance was found to correlate with CN strength. This is a phenomenon also previously observed in 47~Tuc stars \citep{Cottrell1981,Cannon2003}.

The analysis of the high-resolution spectrum of Lee~2525 determined an enhancement in the Zr abundance for this star [Zr/Fe]~$=+0.28$~dex relative to Arcturus. The medium resolution spectra did not reflect this enhancement ([Zr/Fe]~$=+0.02$~dex relative to Arcturus). The mean Zr abundance (relative to Arcturus) for the survey subset showed an enhancement, $\langle$[Zr/Fe]$\rangle =+0.14\pm0.09$~dex, but the large uncertainty is most likely due to the weakness and blending of the measured Zr feature. The mean Ba abundance is $\langle$[Ba/Fe]$\rangle =+0.28\pm0.05$~dex, indicative of a homogeneous distribution of this element in 47 Tuc. The spectral features for La and Nd were deemed to be too blended at this resolution for reliable spectrum synthesis analysis, hence this resolution is too low to carry out a comprehensive survey of {\it s}-process elements in 47 Tuc giant stars.

Most of the light and heavy elements measured in this sample of stars, that all have very similar stellar parameters, have a small spread in values indicating that these elements were most likely created in nuclear processes prior to the formation of these stars. The relation of Na to CN gives the possibility of distinguishing between stellar populations within 47 Tuc. The CN-weak population is typically considered as the initial population, for which the high mass stars have evolved through stellar death and polluted the interstellar medium with products of CNO nulceosynthesis, including the Na enhancement that can be produced through the NeNa cycle. Hence the secondary population are the CN-strong stars. This is consistent with the recent study by \citet{Milone2011}, who used precision photometry from HST and ground-based telescopes to identify multiple stellar populations in 47 Tuc. They note 2 major populations with C, N and Na abundance characteristics similar to those found in our study.

The trend of Na and CN can be viewed as a continuum of Na to CN enhancement. From the baseline of enhancement within the initial CN/Na-weak population the CN-strong stars have been further enriched in CN and Na in relative proportions indicating the same process is responsible (CNO cycling via hot bottom burning). However this process was apparently activated in the CN-strong population, not the CN-weak population. So why do the CN-strong stars have differing degrees of enhancement in CN and Na? Is there another parameter that must be considered? Another consideration is that this sample resides at the connection of the AGB to the RGB. This may provide another distinction with which to understand the spread in CN and Na abundances. 

The Lee~2525 sample results of the 47~Tuc giant stars provide an indication about the abundance patterns observed in the light elements. More detailed analysis of the whole sample of 47 Tuc giants will work with the best stellar atmosphere model for each star and to then determine stellar elemental abundances by spectrum synthesis. The heavy elements will require observations at a higher resolution to obtain better detail in the stellar spectra of these typically weak features.

\section*{Acknowledgments} %If needed
CCW and PLC would both wish to acknowledge the hospitality and financial assistance of the Max Planck Institute for Astrophysics which enabled progress to be made on this paper. PLC would also wish to acknowledge the support of the University of Canterbury for his sabbatical during 2011 and a Marsden Fund grant administered by the Royal Society of New Zealand.

This research has made use of the SIMBAD database, operated at CDS, Strasbourg, France and has made use of data products from the Two Micron All Sky Survey, which is a joint project of the University of Massachusetts and the Infrared Processing and Analysis Center/California Institute of Technology, funded by the National Aeronautics and Space Administration and the National Science Foundation.

\appendix
\section{47 Tuc Giant Star Survey}\label{app:survey}
\onecolumn
\begin{center}
\begin{longtable}{ccccccc}
%\begin{table*}[htbp]
%\begin{minipage}{140mm}
%\multicolumn{7}{c}{\parbox{\LTcapwidth}{Photometry, effective temperature and surface gravity calculated from V-K and CN excess calculated in this study for each of the 97 giant stars in the survey of 47 Tuc.}}
\caption{Photometry, effective temperature and surface gravity calculated from $V$--$K$, and CN excess calculated in this study for the 97 giant stars in our survey of 47 Tuc.} \\
%\end{minipage}
\hline  
Star ID & $V^{*}$ & $B$--$V$ & $V$--$K^{\#}$ & $T_{\textrm{eff}}(V$--$K)$ & $\log g(V$--$K)$ & $\delta$C$_{2011}$ \\ \hline 
\endfirsthead

\multicolumn{7}{c}%
{ \tablename\ \thetable{} -- continued from previous page} \\
\hline Star ID & $V^{*}$ & $B$--$V$ & $V$--$K^{\#}$ & $T_{\textrm{eff}}$($V$--$K$) & $\log g$($V$--$K$) & $\delta$C$_{2011}$ \\ \hline 
\endhead

\hline 
\multicolumn{7}{r}{{Continued on next page}} \\ %\hline
\endfoot

%\hline %\hline
\endlastfoot

%\begin{tabular}{ccccccc}

%Star ID & V$^{*}$ & B-V & V-K$^{#}$ & T$_{eff}(V-K)$ & $\log g(V-K)$ & $\delta$C$_{2011}$ \\ 
Lee4472 & 12.40 & 1.41 & 0.34 & 4274 & 1.44 & -0.112 \\ 
Lee1704 & 13.00 & 1.25 & -0.89 & 4827 & 1.93 & -0.045 \\ 
Pal502 & 12.64 & 1.36 & 3.28 & 4158 & 1.27 & -0.037 \\ 
Lee1628 & 12.94 & 1.22 & 2.99 & 4339 & 1.50 & -0.017 \\ 
Lee1316/Pal881 & 13.17 & 1.17 & 2.89 & 4411 & 1.64 & -0.014 \\ 
Pal1151 & 13.00 & 1.25 & 3.06 & 4295 & 1.50 & -0.011 \\ 
Lee2306/Pal388 & 12.59 & 1.29 & 3.07 & 4288 & 1.33 & -0.005 \\ 
Lee5601 & 12.88 & 1.25 & 3.06 & 4292 & 1.45 & -0.001 \\ 
Lee2428 & 13.23 & 1.16 & 2.96 & 4360 & 1.63 & 0.012 \\ 
Lee3622/Pal726 & 12.60 & 1.31 & 3.24 & 4183 & 1.27 & 0.014 \\ 
Lee1506 & 13.27 & 1.15 & 2.87 & 4423 & 1.69 & 0.016 \\ 
Lee1104/Pal436 & 13.78 & 1.08 & 1.72 & 4714 & 2.07 & 0.017 \\ 
Lee1312/Pal771 & 14.48 & 1.05 & 2.57 & 4927 & 2.46 & 0.021 \\ 
Lee1320/Pal925 & 13.52 & 1.13 & 1.57 & 4509 & 1.84 & 0.022 \\ 
Lee3207/Pal380 & 13.11 & 1.22 & 2.97 & 4352 & 1.58 & 0.028 \\ 
Lee2608 & 12.87 & 1.22 & 3.07 & 4285 & 1.44 & 0.031 \\ 
Lee4737 & 12.96 & 1.25 & 3.10 & 4267 & 1.46 & 0.031 \\ 
Lee4628 & 12.53 & 1.31 & 3.14 & 4241 & 1.28 & 0.043 \\ 
Lee3305/Pal461 & 13.03 & 1.23 & 3.02 & 4319 & 1.53 & 0.044 \\ 
Lee1522 & 13.56 & 1.08 & 1.88 & 4519 & 1.87 & 0.046 \\ 
Lee1423/Pal1124 & 13.75 & 1.03 & 1.68 & 4761 & 2.08 & 0.047 \\ 
Lee4506 & 13.54 & 1.11 & 1.73 & 4454 & 1.82 & 0.050 \\ 
Pal1158 & 13.20 & 1.13 & 2.63 & 4617 & 1.78 & 0.052 \\ 
Lee2616 & 13.21 & 1.14 & 2.84 & 4448 & 1.68 & 0.053 \\ 
Lee2737 & 13.51 & 1.13 & 1.51 & 4553 & 1.87 & 0.053 \\ 
Lee2742 & 13.04 & 1.08 & 2.89 & 4406 & 1.59 & 0.059 \\ 
Lee3404/Pal440 & 14.53 & 0.90 & 2.57 & 4961 & 2.49 & 0.061 \\ 
Lee2311/Pal500 & 13.98 & 0.77 & 2.02 & 5708 & 2.58 & 0.065 \\ 
Lee4502 & 13.28 & 0.96 & -0.51 & 4468 & 1.69 & 0.074 \\ 
Pal256/Tuc14 & 14.00 & 0.85 & 2.06 & 5325 & 2.44 & 0.075 \\ 
Lee2310/Pal423 & 14.50 & 0.94 & 2.46 & 4838 & 2.42 & 0.075 \\ 
Lee1201/Pal507 & 12.88 & 1.26 & 3.10 & 4266 & 1.43 & 0.076 \\ 
Lee1206/Pal638 & 14.13 & 0.84 & 3.37 & 5252 & 2.46 & 0.076 \\ 
Lee5705 & 13.86 & 1.11 & 2.42 & 4505 & 1.98 & 0.077 \\ 
Lee3306/Pal447 & 14.04 & 0.90 & 2.03 & 5288 & 2.44 & 0.079 \\ 
Lee2525 & 12.43 & 1.29 & 3.16 & 4232 & 1.23 & 0.081 \\ 
Chu4432/W66 & 12.55 & 1.34 & 3.24 & 4181 & 1.25 & 0.083 \\ 
Lee1313 & 14.44 & 1.00 & 3.08 & 4855 & 2.40 & 0.084 \\ 
Lee1105 & 13.69 & 1.10 & 2.46 & 4598 & 1.97 & 0.086 \\ 
Lee4514 & 14.35 & 0.98 & 2.34 & 4778 & 2.33 & 0.087 \\ 
Lee1315/Pal869 & 14.10 & 0.81 & 2.85 & 5297 & 2.47 & 0.090 \\ 
Lee3307 & 14.23 & 0.96 & 2.54 & 5448 & 2.58 & 0.092 \\ 
Lee3201 & 13.78 & 0.88 & 1.81 & 5212 & 2.31 & 0.093 \\ 
Lee1529 & 13.90 & 0.86 & 2.80 & 5118 & 2.31 & 0.096 \\ 
Lee8528 & 13.72 & 0.92 & 2.22 & 5031 & 2.20 & 0.105 \\ 
Lee8302/Pal1446 & 14.06 & 0.85 & 1.78 & 5334 & 2.47 & 0.106 \\ 
Pal306/Tuc32 & 14.02 & 0.85 & 2.03 & 5276 & 2.43 & 0.107 \\ 
Lee1516 & 13.94 & 1.04 & 2.22 & 4660 & 2.10 & 0.110 \\ 
Lee3312/Pal466 & 14.11 & 0.88 & 1.88 & 5303 & 2.48 & 0.115 \\ 
Lee1207/Pal5628 & 14.07 & 0.84 & 2.06 & 5262 & 2.44 & 0.127 \\ 
Lee1735 & 13.35 & 1.09 & 2.76 & 4505 & 1.77 & 0.128 \\ 
Pal452 & 12.94 & 1.24 & 2.93 & 4378 & 1.53 & 0.131 \\ 
Lee5703 & 12.53 & 1.28 & 3.17 & 4221 & 1.26 & 0.131 \\ 
Pal1036 & 13.95 & 0.86 & 2.08 & 5341 & 2.43 & 0.131 \\ 
Lee2739 & 13.22 & 1.18 & 2.80 & 4472 & 1.70 & 0.132 \\ 
Lee1513 & 12.41 & 1.32 & 3.14 & 4242 & 1.23 & 0.135 \\ 
Pal487 & 13.17 & 1.15 & 2.87 & 4422 & 1.65 & 0.139 \\ 
Lee3205/Pal351 & 14.07 & 0.83 & 2.05 & 5307 & 2.46 & 0.139 \\ 
Lee2601 & 13.83 & 1.06 & 3.12 & 4638 & 2.04 & 0.144 \\ 
Chu4684W139 & 12.70 & 1.12 & 3.17 & 4223 & 1.33 & 0.146 \\ 
Lee8202/Pal1526 & 14.08 & 0.86 & 3.04 & 5341 & 2.48 & 0.148 \\ 
Lee3415/Pal368b & 14.03 & 0.86 & 1.99 & 5310 & 2.45 & 0.149 \\ 
Lee3403/Pal428 & 14.03 & 0.84 & 2.79 & 5383 & 2.48 & 0.149 \\ 
Lee2108/Pal342 & 14.01 & 0.84 & 2.47 & 5347 & 2.45 & 0.157 \\ 
Lee3310/Pal474 & 14.07 & 0.85 & 2.92 & 5316 & 2.47 & 0.158 \\ 
Lee1301/Pal516 & 12.64 & 1.22 & 2.91 & 4391 & 1.42 & 0.159 \\ 
Lee2604 & 13.07 & 1.04 & 2.52 & 4731 & 1.79 & 0.161 \\ 
Lee8508 & 13.80 & 0.87 & 2.19 & 5006 & 2.22 & 0.162 \\ 
Pal578 & 13.38 & 1.14 & 2.78 & 4488 & 1.77 & 0.164 \\ 
Pal262/Tuc16 & 12.72 & 1.29 & 3.14 & 4240 & 1.35 & 0.166 \\ 
Lee4602 & 13.07 & 1.19 & 2.99 & 4338 & 1.55 & 0.167 \\ 
Lee2201/Pal368a & 13.97 & 0.83 & 1.90 & 5351 & 2.44 & 0.170 \\ 
Chu4241/W164 & 12.62 & 1.12 & 3.05 & 4300 & 1.35 & 0.173 \\ 
Lee3519 & 14.40 & 1.02 & 2.38 & 4755 & 2.34 & 0.174 \\ 
Lee8301/Pal1353 & 13.96 & 0.85 & 2.33 & 5334 & 2.43 & 0.175 \\ 
Pal571 & 13.75 & 1.08 & 1.63 & 4619 & 2.00 & 0.175 \\ 
Lee1319 & 13.08 & 0.93 & 2.28 & 5037 & 1.95 & 0.177 \\ 
Lee1519 & 13.89 & 0.88 & 3.08 & 5299 & 2.39 & 0.181 \\ 
Lee1304/Pal506 & 14.51 & 0.96 & 3.38 & 4794 & 2.40 & 0.185 \\ 
Lee3206/Pal383 & 14.06 & 0.84 & 1.86 & 5303 & 2.46 & 0.186 \\ 
Lee1309 & 13.20 & 1.04 & 2.61 & 4635 & 1.79 & 0.187 \\ 
Lee2419/Pal478 & 14.48 & 0.96 & 2.56 & 4852 & 2.42 & 0.194 \\ 
Lee1324/Pal1125 & 14.04 & 0.85 & 2.13 & 5287 & 2.44 & 0.194 \\ 
Lee4626 & 13.27 & 1.17 & 2.87 & 4422 & 1.69 & 0.196 \\ 
Lee4509 & 12.99 & 1.21 & 3.01 & 4326 & 1.51 & 0.198 \\ 
Lee1408/Pal598 & 13.49 & 1.09 & 2.69 & 4566 & 1.87 & 0.201 \\ 
Lee2511 & 13.84 & 0.84 & 1.94 & 5202 & 2.33 & 0.201 \\ 
Pal661 & 12.54 & 1.26 & 3.15 & 4233 & 1.28 & 0.202 \\ 
Lee1419 & 14.19 & 0.97 & 3.37 & 4795 & 2.27 & 0.208 \\ 
Lee5717 & 13.16 & 1.16 & 2.94 & 4372 & 1.61 & 0.209 \\ 
Lee5514 & 14.15 & 1.00 & 2.21 & 4705 & 2.21 & 0.216 \\ 
Lee1747/S364 & 12.45 & 1.26 & 3.07 & 4284 & 1.27 & 0.223 \\ 
Lee4636 & 13.30 & 1.17 & 2.78 & 4493 & 1.75 & 0.233 \\ 
Lee3510 & 13.63 & 1.09 & 2.82 & 4780 & 2.04 & 0.235 \\ 
Lee4515 & 14.49 & 0.98 & 2.45 & 4848 & 2.42 & 0.262 \\ 
Lee2302/Pal397 & 14.07 & 1.07 & 2.12 & 4720 & 2.19 & 0.264 \\ 
Lee2528/Pal625 & 13.54 & 1.12 & 1.52 & 4532 & 1.87 & 0.307 \\ 
 &  &  &  &  &  &  \\ 
 & \multicolumn{3}{l}{$^*$ $B$\&$V$ from SIMBAD}   &  &  &  \\ 
 & \multicolumn{3}{l}{$^\#$ $K$ from 2MASS}   &  &  &  \\ 
%\end{tabular}
\hline
\label{all97_info}
%\end{table*}
\end{longtable}
\end{center}
\twocolumn

\end{document}